\documentclass[journal]{vgtc}                
\ifpdf
  \pdfoutput=1\relax                   
  \usepackage{graphicx}                
  \DeclareGraphicsExtensions{.pdf,.png,.jpg,.jpeg} 
\else
  \usepackage{graphicx}                
  \DeclareGraphicsExtensions{.eps}     
\fi%

\graphicspath{{figures/}{pictures/}{images/}{./}} 

\usepackage{microtype}                 
\PassOptionsToPackage{warn}{textcomp}  
\usepackage{textcomp}                  
\usepackage{mathptmx}                  
\usepackage{times}                     
\usepackage{cite}                      
\usepackage{tabu}                      
\usepackage{booktabs}                  
\usepackage{balance}

\usepackage{soul} 

\usepackage{xcolor,colortbl} 
\usepackage[hyphens]{url}
\renewcommand\hl[1]{#1}
\usepackage{amssymb} 

\onlineid{1025}

\vgtccategory{Research}
\vgtcpapertype{evaluation}

\title{Quantifying the Effects of Working in VR for One Week}

\newcommand*\samethanks[1][\value{footnote}]{\footnotemark[#1]}
\author{Verena Biener\thanks{Coburg University of Applied Sciences, Germany}, Snehanjali Kalamkar\samethanks, Negar Nouri\samethanks, Eyal Ofek\thanks{Microsoft Research}, Michel Pahud\samethanks, John J. Dudley\thanks{University of Cambridge}, Jinghui Hu\samethanks, \\Per Ola Kristensson\samethanks, Maheshya Weerasinghe\thanks{University of Primorska, Slovenia}, Klen Čopič Pucihar\samethanks, Matjaž Kljun\samethanks, Stephan Streuber\samethanks[1], Jens Grubert\samethanks[1]}

\abstract{Virtual Reality (VR) provides new possibilities for modern knowledge work.  
However, the potential advantages of virtual work environments can only be used if it is feasible to work in them for an extended period of time. Until now, there are limited studies of long-term effects when working in VR. This paper addresses the need for understanding such long-term effects. Specifically, we report on a comparative study ($n$=16), in which participants were working in VR for an entire week---for five days, eight hours each day---as well as in a baseline physical desktop environment. 
This study aims to quantify the effects of exchanging a desktop-based work environment with a VR-based environment. Hence, during this study, we do not present the participants with the best possible VR system but rather a setup delivering a comparable experience to working in the physical desktop environment.
The study reveals that, as expected, VR results in significantly worse ratings across most measures. Among other results, we found concerning levels of simulator sickness, below average usability ratings \hl{and two participants dropped out on the first day using VR, due to migraine, nausea and anxiety}.
Nevertheless, there is some indication that participants gradually overcame negative first impressions and initial discomfort.
Overall, this study helps lay the groundwork for subsequent research, by clearly highlighting current shortcomings and identifying opportunities for improving the experience of working in VR. 
} 

\keywords{virtual reality, long-term, knowledge work, user study}

\teaser{
  \centering
  \includegraphics[width=\linewidth]{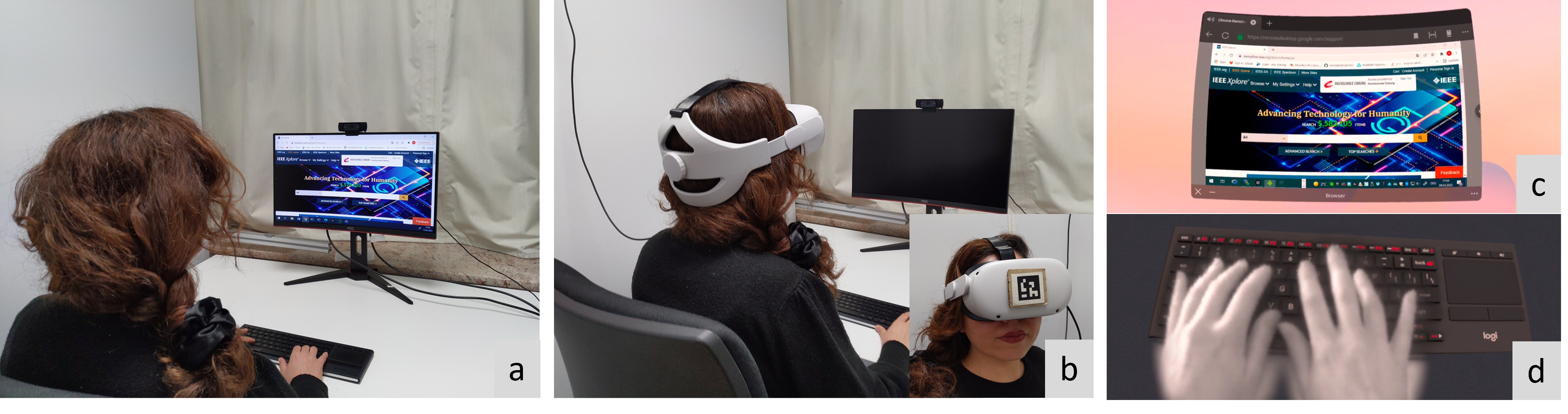}
  \caption{a) Participant working in a physical desktop setup in one of three physical locations (Germany), including a curved display and a keyboard with integrated touchpad. b) Participant working in the VR setup wearing an Oculus Quest 2, using the same keyboard but a virtual curved display. c) Participant's view within VR where the streamed content from a remote machine is visible. d) Participant's view of the keyboard and hands within VR.}
  \label{fig:teaser-setup}
}

\vgtcinsertpkg

\begin{document}


\firstsection{Introduction}

\maketitle


Virtual Reality (VR) has the potential to enhance physical working environments, for instance, by providing repeatable, location independent user experiences, or relieving physical world limitations such as screen sizes of physical screens~\cite{grubert2018office}. For example, prior work studied transplanting or extending tasks performed by knowledge workers (as defined by \cite{drucker1966effective})
from physical 2D displays to head-mounted displays (HMDs) (e.g.~\cite{Biener2020Breaking, gesslein2020pen, pavanatto2021we}). The use of a large display space around the user provided by the HMD, not limited by the size of physical monitors, supports visualization of information in multiple depth layers. The direct manipulation of data using natural hand motions, and the ability to map small physical motions to larger actions in such virtual environments can reduce fatigue and may open up the workspace for people with special needs. Remote collaboration in such environments brings people to the same virtual space, and, with a varying level of representation needed for a particular task at hand, can increase their sense of presence \cite{pejsa2016Room2Room}. In addition, an office in VR can dynamically adapt to a user's work situation---it could transform into a calming beach when reading a paper or a formal office when writing an email. Virtual environments also provide privacy from the outside world and the removal of real environment disruptions may help users focus their attention on work. Further, collaborative work in virtual environments can save users the travel hassle, associated costs and reduce carbon footprint~\cite{jordan2020virtual}.

However, VR substitutes users' visual, audio and sometimes haptic sensations provided by the physical world with artificial inputs. With current VR technologies many of these inputs provide an inferior experience compared to the real world.
For example, HMDs typically have a smaller field of view (FoV) compared to humans' visual field, and their resolution, while increasing over the years, is still lower than the retinal resolution. Most HMDs render the entire virtual world at a fixed focal distance and their dynamic range is smaller than what we can perceive as humans. Additionally, with their substantial weight and by blocking the air flow on the face, HMDs can reduce users' comfort.
While many of these limitations are likely to be addressed and improved in the future, visions of virtual knowledge work are not limited to research labs anymore. Startups such as Spatial or large corporations such as Meta and Microsoft are advancing products to facilitate virtual knowledge work.

Hence, we see it as beneficial to understand \emph{how} VR technology available to end-users today influences knowledge work.  In particular, we strive to quantify the effects of VR experiences that users can have today using commercial of-the-shelf hardware. Hence, we decided against designing an as good as possible virtual environment. Instead, we intentionally decided to study a specific operating point of VR hardware and software---the commercially available Oculus Quest 2 HMD and Logitech K830 physical keyboard with integrated touchpad. This specific combination enables integrated hand and keyboard tracking and allows virtual work for a large user group today.

To this end, we report on a study with users working for an entire workweek (five days, eight hours each day) in VR with the aforementioned setup (\autoref{fig:teaser-setup}, b). To quantify the costs of this setup compared to physical workplaces, participants also worked another workweek in a corresponding physical work environment (\autoref{fig:teaser-setup}, a). To be able to compare the two conditions, we made both the virtual and physical work environments as similar as possible: the same display size, shape, resolution and input device. At the same time, and in contrast to prior studies (e.g.~\cite{ruvimova2020transport, guo2019mixed}), we decided against prescribing artificial tasks to participants but instead allowed participants to determine their own actual work. While this choice potentially impacts repeatability and replicability it also increases the ecological validity of the findings.






We are well aware that with the current state of VR technology, working in VR will be demanding on the user: 
the size, weight and quality of the HMD, its limited FoV, latency, and the authenticity of the representation of the world around users (drinks, keyboard, mouse, etc.) can affect the workflow. 
As a result, one can expect that the user experience in VR might be inferior to the one in the physical environment. Still, we see it as a worthwhile endeavor to \emph{quantify} the effects of working in VR over five consecutive working days with eight hours in each day. This can serve as baseline for future optimized VR experiences that do not necessarily replicate a physical environment. Further, running a study over five consecutive days allows us to study gradual changes over time. 

The main findings of the study are as follows: 1) self-rated task load was significantly higher in VR (approximately 35\%), as was frustration (42\%), negative affect (11\%), anxiety (19\%) and eye strain (48\%); 2) VR resulted in significantly lower system usability scale scores (36\%) with below average ratings, self-rated flow (14\%), perceived productivity (16\%) and wellbeing (20\%); 3) VR resulted in (according to Stanney et al.~\cite{stanney1997cybersickness}) poor ratings of simulator sickness. \hl{The values of some measures improved over the five days} for both \textsc{vr} and \textsc{physical}. However, we only found that the rate of change was significantly higher for \textsc{vr} regarding visual fatigue ratings compared to \textsc{physical} which means it decreased significantly faster in \textsc{vr}.

In order to facilitate future work, we release an anonymized dataset along with this paper \hl{which can be found at} \url{https://gitlab.com/mixedrealitylab/quantifying-the-effects-of-working-in-vr-for-one-week}. 
In summary, this paper presents the following contributions: 1) a study on the effects of working in VR for five working days, eight hours per day, 2) quantification of effects that occur between VR and a baseline physical work environment and 3) an accompanying dataset to aid replication and further analysis by the community.

\section{Related Work}
There are several potential benefits of using VR as a working environment. Besides theoretical advantages of virtual offices (e.g.,~\cite{grubert2018office, ofek2020towards, knierim2020virtual, knierim2021nomadic} further work has empirically investigated specific benefits of VR for knowledge work, which are discussed in  \autoref{section:benefitsOfVR}. 
Yet, the prolonged usage of VR can affect the usage and workflow, and 
the following (\autoref{section:workingForLongTime}) discusses previous work investigating long-term VR use. The potential benefits and understanding long-term usage motivated our work, which builds upon and extends the existing body of literature. 

\subsection{Benefits of VR for Knowledge Work}
\label{section:benefitsOfVR}
Prior research has examined the use of VR as an environment for knowledge work, showing a variety of effects of VR on the quality of work, the flexibility of the VR display, the direct interaction which can be used to increase productivity and the control of the environment around the worker which can be used to reduce stress.
Ruvimova et al.~\cite{ruvimova2020transport} found that a VR office on a virtual beach successfully reduced distraction and simulated the workflow of a closed physical office. This indicates that VR can help users stay focused in open office environments. 
This capability is not restricted to VR. By displaying virtual separators in a physical office, augmented reality (AR) has also been shown to be useful against visual distractions~\cite{lee2019partitioning}.  
In addition, users were allowed to personalize their work environment, which helped to increase their satisfaction and improved their experience of a shared workspace. Personalization of workspace has been shown beneficial in several studies and the literature emphasizes the importance of personalized design concepts for the so called non‐territorial offices or shared office space~\cite{brunia2009personalization, kim2016desk}. In such contexts, both VR and AR have the capability to allow users to design and decorate their virtual office space according to their preferences. 

Prior research~\cite{anderson2017relaxation, thoondee2017using} has demonstrated how virtual nature environments can reduce stress and improve mood during work. It is well known that spaces filled with greenery or even a view on greenery provide an opportunity for recovery from mental fatigue and are generally beneficial to human health~\cite{berman2008cognitive, kaplan1995restorative}. Further work indicates that VR can reduce stress more effectively than simply streaming a video of relaxing content~\cite{pretsch2020improving} and that interactive VR environments are more effective than passively consuming VR content~\cite{valtchanov2010restorative}. Also, Mostajeran et al. \cite{mostajeran2021effects} found that showing a forest environment had positive effects on cognition. Despite these benefits, Li et al.~\cite{li2021rear} found that while users preferred nature environments, they were more productive when working in an office-like environment. 






VR also provides new possibilities for interacting with and visualizing work-related content. Biener et al.~\cite{Biener2020Breaking} showed how multimodal interaction techniques, including eye tracking, can be used to efficiently navigate between a large number of virtual displays and how a three-dimensional visualization with depth-perception makes tasks involving multiple layers much easier. Pavanatto et al.~\cite{pavanatto2021we} compared physical and virtual monitors in AR and concluded that while virtual monitors can be beneficial, they are still technically inferior. Therefore, they suggested to combine both physical and virtual monitors in the workflow, which is already put to practice in commercial products such as the Lenovo ThinkReality A3 Smart Glasses. 

VR can also increase the usability and performance of knowledge worker tasks. For example, spreadsheet~\cite{gesslein2020pen} and presentation authoring applications~\cite{biener2022povrpoint} can significantly benefit from depth perception, potentially large virtual display space, and new interaction possibilities such as a spatially tracked pen and eye-tracking, all provided by VR as well. 

The benefits of extended virtual display spaces have been investigated in both  stationary~\cite{mcgill2020expanding, ens2014personal} and mobile~\cite{ng2021passenger, menzner2020above} environments. For example, new interaction techniques for navigating  large display spaces have been proposed using non-linear mapping of head-gaze~\cite{mcgill2020expanding} or above-surface interaction~\cite{menzner2020above}. Personalized layouts of multiple virtual displays have also been studied~\cite{ens2014personal, ng2021passenger}. For example, Ens et al.~\cite{ens2014personal} indicated that application switching times can be decreased by up to 40\% using optimized layouts. Ng et al.~\cite{ng2021passenger} found that layout preferences depend on the perception of other passengers' physical presence, e.g., when sitting in an airplane seat next to another person.
Using VR for knowledge work can also address privacy issues~\cite{grubert2018office, schneider2019reconviguration} and enhance capabilities of existing devices, such as changing the keyboard on the fly to support other languages, symbols, and layouts~\cite{schneider2019reconviguration}.

However, these prior works have been studied in short term experiments. Also, while the proposed interaction techniques show the potential for supporting knowledge work in VR they often are restricted to specific lab-based setups. Instead, in our study, we focused on experiences that are accessible to potentially millions of users today, relying on commercially off-the-shelf hardware and software.

\subsection{Working in VR for an Extended Period of Time}
\label{section:workingForLongTime}
While several possibilities of using VR for knowledge work have been explored in recent years, showing advantages, the effects of long-term usage of such environments are still not sufficiently understood. Besides anecdotal reports (for example \cite{buckley2017two,klemens2022Ispent}), 
researchers began to investigate long-term effects of AR and VR interfaces. 
Long-term usage of AR has been explored in manufacturing. In a study by Grubert et al.~\cite{grubert2010extended} participants used AR for four hours in an order picking task. The results showed a higher work efficiency in the AR condition without an increase in overall objective and subjective strain. However, some participants felt a higher eye discomfort compared to a non-AR baseline.
Wille et al.~\cite{wille2014prolonged} compared four hours of work on a monocular HMD, tablet and computer monitor, and did not find any objective physiological effect on participants' visual system. However, subjective ratings of strain were significantly higher for the HMD, which authors attributed to the unfamiliarity with the technology. 
In Funk et al.'s study~\cite{funk2017working}, participants used a projector-based AR system for at least three full working days to perform an assembly task. The system was shown to be useful for untrained workers, yet it slowed down the performance and increased cognitive load for expert workers. 

Lu et al.~\cite{lu2021evaluating} conducted an in-the-wild study of glanceable AR for everyday use where participants used the prototype for three days. The authors concluded that participants liked the prototype and would use it daily if the HMD had a form factor of regular prescription glasses. 
In a 24-hour self-experiment~\cite{steinicke2014self} one participant worked, ate, slept and entertained himself in VR. 
In another case study~\cite{nordahl201912}, two participants used VR HMDs for 12 hours straight and reported only mild simulator sickness symptoms.






The prior work most similar to this paper has been conducted by Guo et al.~\cite{guo2019mixed, guo2019evaluation, guo2020exploring} and Shen et al.~\cite{shen2019mental}. They looked at prolonged use of VR for office work in a virtual environment. 
Their long-term study~\cite{guo2019mixed, guo2020exploring, shen2019mental} had 27 participants working in a virtual and a physical office for eight hours, performing tasks such as document correction, keyword searching, text input and image classification. 
In the context of the Maslow's Hierarchy of Needs, the researchers found that emotional needs must be met in short and long-term use, while physiological, belongingness needs, temporal- and self-presence are only important for long-term use~\cite{guo2019mixed}. 
Evaluating the effects on visual discomfort, Guo et al.~\cite{guo2020exploring} found that signs of visual fatigue (subjective rating, pupil size, accommodation response) change with time in both physical and virtual conditions. They also indicated that female participants suffered more from visual fatigue and they speculated that this could be due to less experience with VR. They did not find significant differences in nausea and eye strain between the virtual and physical condition, but larger difficulty in focusing and physical discomfort (due to weight and form factor of HMD) were present in VR.
In addition, Shen et al.~\cite{shen2019mental} asked participants to perform a psychomotor vigilance task (PVT) six times during the day and found significantly less PVT lapses and higher reaction times in the physical environment, which indicates a higher mental fatigue in VR. They propose two explanations: either VR occupies more attentional resources or VR can increase attention of participants more effectively so that participants use more attentional resources. However, all these findings are based on artificial tasks designed by researchers, which might influence the results compared to real in-the-wild tasks. Also, prolonged use studies seem necessary to ensure participants obtain sufficient familiarity with the VR setup so that it is comparable with working in the physical environment. Hence, we here conduct a study that is five times as long in duration (five days, eight hours per day in each condition) compared to the prior studies 
\cite{guo2019mixed, guo2020exploring, shen2019mental}. 
Further, in our study we ask participants to work on their own everyday work tasks.

\section{Study}
Our goal was to quantify the effects of working in a virtual reality environment for extended periods of time. Specifically, we compared working in VR to working in a physical environment by analyzing their respective effects on a variety of measures as explained in the following sections. The study was approved by \hl{the ethics committee of Coburg University of Applied Sciences and Arts}.

\subsection{Study Design}
The experiment was designed as a within-subjects study. Each participant completed a full week of work (5 days, 8 hours per day) in a virtual and in a physical environment. The first independent variable is \textsc{environment}, which is either \textsc{vr} or \textsc{physical}. \hl{We counterbalanced \textsc{environment}, with half the participants starting with the \textsc{vr} week and half of them with PHYSICAL.}
The second independent variable is \textsc{day} with five levels, namely \textsc{day1}, \textsc{day2}, \textsc{day3}, \textsc{day4} and \textsc{day5}.

To get a better idea of how the measures evolve during the week, participants answered various questionnaires five times a day \hl{at fixed times}: before starting the daily work, two before lunch break \hl{(after 2 and 4 hours of work)} and two after \hl{a 45 minute} lunch break \hl{after 6 and 8 hours of work}. We subsequently merged these data-points into three values for each day, resulting in a third independent variable \textsc{time} with the three levels \textsc{start}, \textsc{morning} (mean of two data points before lunch break) and \textsc{afternoon} (mean of two data points after lunch break).
\hl{We merged the two data points from morning and afternoon to make measures more robust against the influence of different work tasks and to account for possible logging errors, as one missing value would eliminate all measures from this participant for ANOVA analysis.
}

Dependent variables included self-rated subjective as well as objective measures as explained in the next section. The number of data points, and, therefore, also the number of levels of the independent variables \textsc{day} and \textsc{time} varies between these measures, because depending on their purpose, some of them were recorded at different frequencies. The measures and the number of data-points obtained for each are summarized in \autoref{tab:measurements}. 


\begin{table}[t]
    \centering 
    \caption{Overview of measures taken during the study: before starting the daily work (\textsc{start}), and after two (2), four (4), six (6) and eight (8) hours of work. For the significance tests the average of each of the two values from the morning and afternoon are used. }
    \small
    \begingroup
    \setlength{\tabcolsep}{5pt}

    \begin{tabular}{|>{\centering}p{1.9cm}|>{\centering}p{0.9cm}|>{\centering}p{0.9cm}|>{\centering}p{0.9cm}|>{\centering}p{0.9cm}|>{\centering\arraybackslash}p{0.9cm}|}
        \hline
        Measure & Start & 2h & 4h & 6h & 8h    \\ 
        \hline 

        Task Load &   & \checkmark & \checkmark & \checkmark & \checkmark   \\
        \hline 
        
        System Usability &  & \checkmark & \checkmark & \checkmark & \checkmark   \\ 
        \hline 

        Flow &   & \checkmark & \checkmark  &  \checkmark & \checkmark    \\
        \hline 

        Perceived Productivity &  & \checkmark & \checkmark  &  \checkmark & \checkmark   \\
        \hline 

        Frustration &   & \checkmark & \checkmark &  \checkmark & \checkmark   \\
        \hline
        
        Presence &  & \textsc{vr} only &  \textsc{vr} only & \textsc{vr} only &  \textsc{vr} only   \\
        \hline 
        
        Pos. / Neg. Affect & \checkmark & & &  & \checkmark  \\ 
        \hline
        
        Wellbeing & \checkmark & \checkmark & \checkmark & \checkmark & \checkmark  \\
        \hline 
        
        Anxiety & \checkmark & \checkmark & \checkmark & \checkmark & \checkmark  \\
        \hline 

        Simulator Sickness & \checkmark & \checkmark & \checkmark & \checkmark & \checkmark  \\
        \hline 

        Visual Fatigue & \checkmark & \checkmark & \checkmark & \checkmark & \checkmark   \\
        \hline 
        
        Heart Rate & \multicolumn{5}{c|}{continuous}   \\
        \hline 
        
        Break Times & \multicolumn{5}{c|}{continuous}  \\
        \hline 
        
        Typing Speed & first day  &  & & & last day   \\ 
        \hline 
    \end{tabular}

    \endgroup

    \label{tab:measurements}
\end{table}

\subsection{Measures}
\label{sec:Measures}
In the course of the study we collected a range of subjective as well as objective measures
which are presented in \autoref{tab:measurements}. Some of the subjective measures were assessed with questionnaires at four specified time intervals: after two, four, six and eight hours of work. These measures include:
task load measured by the NASA TLX questionnaire~\cite{hart1988development}, usability measured by the system usability scale~\cite{brooke1996sus}, flow~\cite{rheinberg2003erfassung}, presence using the IPQ questionnaire\footnote{\url{http://www.igroup.org/pq/ipq}} \cite{schubert2003sense}, and two separate questions asking the participants to rate their \hl{perceived} productivity and frustration on a 7-point-Likert-scale (from 1 to 7) (``I was very productive in the last two hours.''; ``I was very frustrated in the last two hours.''). 
These measures where 
not taken at the \textsc{start} of the day, because they are not meaningful without referring to a prior period of work. 


Further subjective measures were assessed five times a day. Specifically, data was collected before starting the daily work as well as every two hours as with the aforementioned questionnaires. These measures assessed different aspects of participants' general physical and mental wellbeing: anxiety using the short version of STAI~\cite{zsido2020development} to examine if \textsc{vr} has an effect on anxiety, the simulator sickness questionnaire~\cite{kennedy1993simulator}, visual fatigue using six questions from~\cite{heuer1989rest} as was done by~\cite{benedetto2013readers}, and a separate question asking participants to rate their wellbeing on a 7-point-Likert-scale (``I was very comfortable in the last two hours.''). Even though the simulator sickness questionnaire already includes one question about visual fatigue, we chose to use an additional set of questions to assess it more closely.
In addition, participants answered the PANAS-SF questionnaire~\cite{watson1988development} to record a positive/negative affect of each condition on emotional state of users. This questionnaire was answered solely before starting the daily work and at the end of the working day to measure a change in emotional state induced by a the whole day of work. Sometimes, participants could not complete a particular questionnaire at the exact time because they had to participate in a meeting or lecture. Still the average duration of time blocks was very close to two hours ($m=122~min$, $sd=55.88$)

At the end of each week, further qualitative data was collected in a short interview. Participants were asked to talk about what they liked or disliked during the week, how they felt, what problems occurred and what they would improve in the particular condition used that week. After both weeks were completed, they were also asked about their preferences and if they could imagine using VR for work in the future.

Additionally, we also collected a set of objective measures. These included heart rate, which can be used as an indication for stress. It was continuously recorded using a Polar H10 chest strap. Using a webcam, we also recorded users' heads and we used these videos to track participants' break behaviors, that is, how much time participants spent away from the screen. Longer breaks were also recorded by ManicTime \cite{manictime2022} software, which we used to additionally detect and confirm inactivity of more than 10 minutes as a break. At the beginning and end of each week the typing speed of each participant was assessed using a web-based typing speed test \cite{10fastfingers2022} to see if they were adapting to the unknown keyboard during the week. Note that we did not focus on text entry performance as a primary measure, as prior work already indicated that users can adapt well to physical keyboards in VR~\cite{grubert2018text, pham2019hawkey}.

\subsection{Apparatus}
The experimental setup was designed to make the work environment in both conditions (\textsc{vr} and \textsc{physical}) comparable, while still relying on commercial off-the-shelf hardware and standard system software. We used an Oculus Quest 2 HMD (Quest-Build 37.0) as it provides integrated tracking of user's hands and a physical keyboard, which was, at the time of the experiment, the Logitech K830. This keyboard has an integrated touchpad and was used as the main input device. Hence, we restrained from using an external mouse, which would also inhibit the repeatability of the experiment (as custom solutions would be needed for mouse tracking).
In both conditions participants used a \textit{work-computer} to work on during the whole experiment. Participants 
could either bring their own computer, or use the computer provided by us. 
In both conditions, a browser and Chrome Remote Desktop \cite{chromeremote2022} were used to connect to the \textit{work-computer}. This allowed participants to see the desktop environment through the Oculus Browser in VR HMD.
To make the \textsc{physical} condition as similar as possible to \textsc{vr} and to reduce confounding variables, we used a second computer (\textit{display-computer}) in the non-VR condition as well 
and connected it to the \textit{work-computer} via Chrome Remote Desktop. 
\hl{Therefore, using a personal laptop did not affect the study as users were accessing it remotely, solely using the curved monitor or the VR HMD and the keyboard.}
The language settings of the \textit{work-computer} were set to English regardless of the geographical location of the participants, because at the time of the study, the Oculus Quest 2 only supported visualization of the English keyboard layout.
A 24-inch curved display (AOC Gaming C24G1) was used as the \textit{display-computer}, which was placed 60~cm from the participants, to resemble the field of view of the virtual browser window in VR (ca. 47\textdegree~horizontally, 27.5\textdegree~vertically) as closely as possible.
The resolution of the \textit{work-computer} was set to 1366$\times$768 at 125\% display scaling. This resolution allowed common user interface elements to still be legible in the virtual display. For example, the capital letter \textit{A} rendered in the typeface Calibri at size 12 pt would result in a vertical FoV of 17.19 arcminutes at 60~cm viewing distance.

The physical curved display was present in both conditions, because a webcam was mounted on top of it to detect the participants' head movements. For the VR condition, an ArUco marker \cite{garrido2014automatic} was attached to the headset to allow analysis of break times. For the non-VR condition, we used the face-mesh algorithm of mediapipe \cite{mediapipe2022}. 
In both conditions, the keyboard was connected directly to the \textit{work-computer} via the Bluetooth, as otherwise the remote connection causes some keys to function incorrectly.  This necessitated the use of a second keyboard in the VR condition which was connected to the Oculus Quest 2 via Bluetooth. When the Quest 2 has a Bluetooth connection to a K830 keyboard, it will detect and display any K830 in view. This allowed the keyboard connected to the \textit{work-computer} and used by the participant to be displayed in VR. The keyboard actually connected to the Quest 2 was hidden out of sight.

In conjunction with the keyboard tracking in the Oculus Quest 2, the hand tracking was also enabled. We  added a virtual desk using the `Bring Your Desk Into VR' option of the Oculus Quest 2. To mitigate distractions in the virtual environment, we selected the `Bubbles' home environment. We did not replicate the rest of the physical environment (e.g.,~walls surrounding the user). While this might create a potential confound, we decided on such design in favor of repeatability and ecological validity of the experiment (i.e. experiences that actual users can have outside of lab environments without the need to install custom hardware or software). Also, while we did not quantify the illuminance at the user's eye in the physical workplace we ensured that the perceived lighting conditions were comparable. 

\begin{figure}[t]
	\centering 
	\includegraphics[width=1\columnwidth]{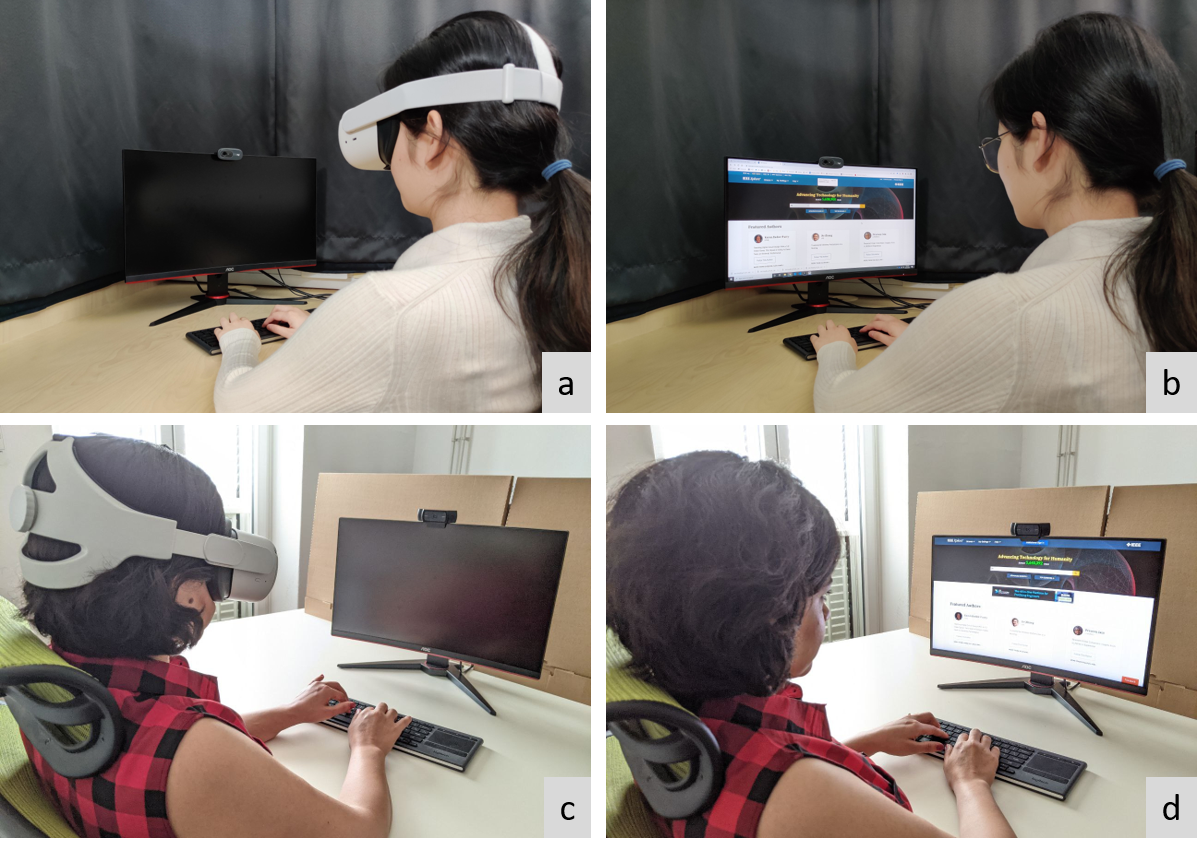}
	\caption{a) \textsc{vr} condition in the UK, b) \textsc{physical} condition in the UK, c) \textsc{vr} condition in Slovenia, d) \textsc{physical} condition in Slovenia.}
	\label{fig:differentSetups}
\end{figure}

In addition, participants could double-tap the side of the HMD to toggle the pass-through-mode, so that they could more easily drink, eat or pick up their phone. 
We also reduced distractions in the physical environment as much as possible, however, the setup slightly differed between the three physical locations in which the experiment was carried out (Germany, UK, Slovenia). In Germany the participants (ten in total) were partially shielded from the rest of the room using mobile walls, as can be seen in \autoref{fig:teaser-setup}, a and \autoref{fig:teaser-setup}, b. In the UK, participants (2) were sitting on their own in the corner of a vacant open office, as displayed in \autoref{fig:differentSetups}, a and \autoref{fig:differentSetups}, b. In Slovenia, participants (4) were sitting in a corner of a small office, 
as shown in \autoref{fig:differentSetups}, c and \autoref{fig:differentSetups}, d. In general, participants were sitting on their own, but due to the length of the experiment we could not completely control the occasional presence of other people. In these situations, other people present were asked to be as quiet as possible. 

In both conditions participants were wearing a Polar H10 chest strap
, which was used for collecting heart rate data. The strap sent data to an Android Phone with the Polar App installed via Bluetooth. 


\subsection{Participants}
In total, 16 participants ($mean~age = 29.31$, $sd = 5.52$, 10 male, 6 female) participated in the study and completed both weeks. All participants were employees or researchers at a university. Two additional participants (age 32 and 33, one male, one female) dropped out on the first day of the \textsc{vr} condition due to a migraine, nausea and anxiety. Eight participants started the study with the \textsc{vr} week and the other eight with the \textsc{physical} week.
Three participants were left handed, but all participants used their right hand to operate a mouse which was consistent with the touch pad of the K830 being on the right-hand side of the keyboard.
All participants had normal, or corrected to normal eyesight and they saw everything clearly in the virtual environment. 
Two (2) participant had no previous experience with VR, six (6) only slight experience, two (2) had moderate experience, four (4) substantial experience, and two (2) extensive experience.
Participants were also asked to indicate how often they usually look at the keyboard while typing on a scale from 1 (never) to 7 (all the time), which resulted in a mean rating of $3.19$ ($sd=1.38$). 
When asked about how often they use a touchpad the mean rating (on the same scale) was $2.88$ ($sd=1.67$). 

\subsection{Procedure}


All participants were informed about the procedure and the content of the study, signed a consent form and filled out a demographic questionnaire.
Next, the participant attached the Polar H10 heart rate sensor. 
On the first day of VR, participants received a short introduction using the HMD, how to activate the pass-through mode, and how to reconnect via Chrome Remote Desktop.

Next, the camera recording of the participant's face and ManicTime software were started and the 
participant filled out the first set of questionnaires (the further ones followed \hl{after 2, 4, 6 and 8 hours} according to \autoref{tab:measurements}). 
\hl{The questionnaires were filled out on the same screens that the participants worked on (a physical screen in \textsc{physical} and virtual one in \textsc{vr}).}
Each participant conducted a typing speed test before the first day and following the last day of each condition as explained in \autoref{sec:Measures} and visible in \autoref{tab:measurements}. 
All participants were required to take a 45 minute break after four hours. All together, the duration of the whole workday was 8 hours 45 minutes. 
At the end of each week participants were interviewed and all data collected during the week was secured.


\subsection{Results}
We used a three-way repeated measures analysis of variance (ANOVA) to analyze the collected data.
Non-normal data was log-transformed \hl{(heart rate, break time, typing speed)} and for the subjective feedback from questionnaires we used the Aligned Rank Transform~\cite{wobbrock2011aligned} before conducting ANOVA \hl{(task load, system usability, flow, perceived productivity, frustration, presence, positive/negative affect, wellbeing, anxiety, simulator sickness, visual fatigue)}.
For multiple comparisons in post hoc tests, \hl{we used Wilcoxon signed-rank test for ART data and t-test otherwise,} both with Bonferroni adjustments at an initial significance level of $\alpha=0.05$.
As already mentioned, the number of independent variables and the number of levels of each independent variable differ between the measures. The results for the measures are presented in the following sections. We only display main effects of the \textsc{environment} and interaction effects involving the \textsc{environment}, \hl{as our focus was on exploring the differences between VR and PHYSICAL and not general variances over time. Therefore, no main effects of DAY or TIME are reported in the paper. We provide a more extensive analysis of other main and interaction effects in the supplementary material.} 
\hl{In addition to examining interaction effects of DAY and \textsc{environment}, we compared the slopes of a fitted line through all days between \textsc{vr} and \textsc{physical} by using a one-sided t-test. We chose a one-sided t-test, because we wanted to know if the slopes for \textsc{vr} are significantly higher, as we hypothesize that \textsc{vr} changes are greater, as participants are getting used to a relatively new system while the \textsc{physical} environment is familiar from the start.}
We only report this, for the measures with interaction effects between \textsc{environment} and \textsc{day} (negative affect, anxiety, simulator sickness, visual fatigue).
\hl{Due to data logging errors, we had to remove some participants from the analysis of several measures (as indicated below). Data logging errors occurred due to the following reasons: 1) the website providing the questionnaires was once not reachable; 2) participants answered the wrong questionnaire-set four times (after 8 hours they did the questionnaire meant for after 6 hours, which did not include positive/negative affect); 3) one participant skipped the first questionnaire once; 4) Polar's heart rate logging application failed to properly sync data for 6 participants on at least one day. As data logging errors rarely occurred for questionnaires (5 times), we do not believe it affected the results.}

\begin{figure*}[tb]
	\centering 
	\includegraphics[width=2\columnwidth]{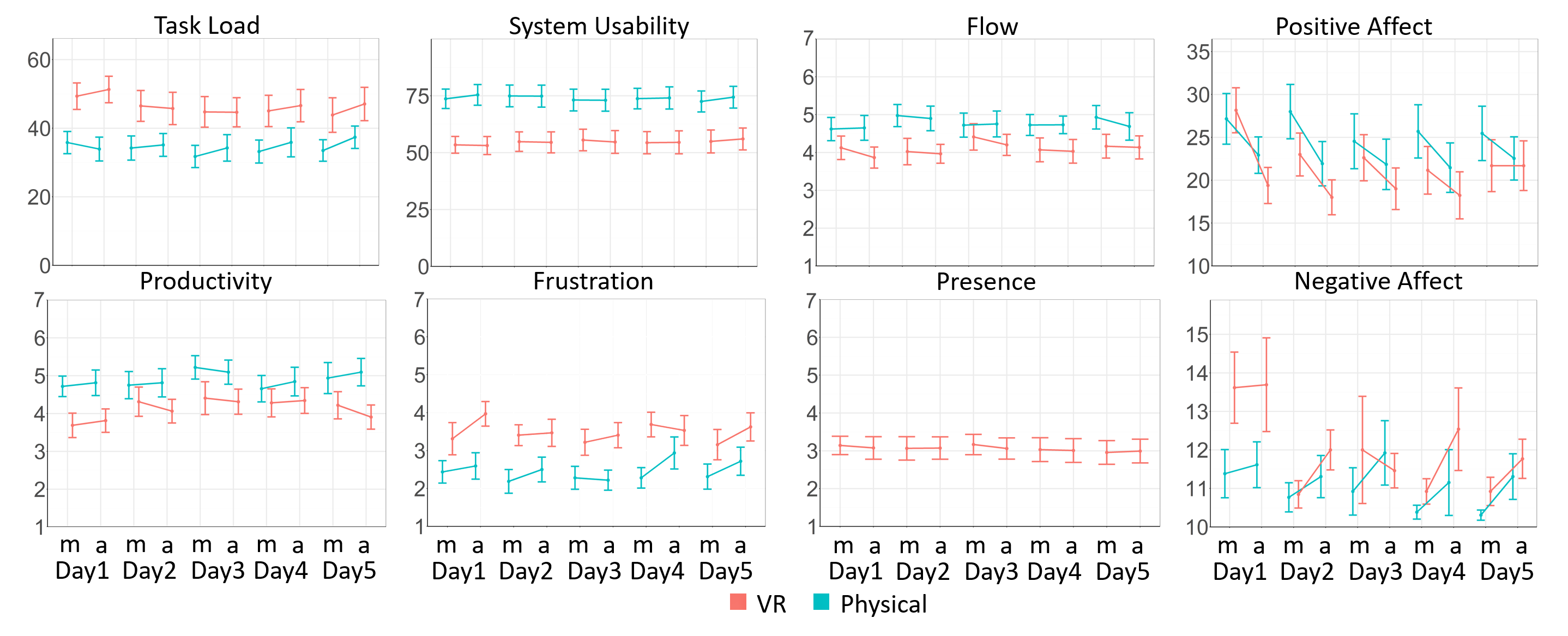}
	\caption{Average values and standard error for subjective measures in the morning (m) and afternoon (a).}
	\label{fig:allPlots_m_a}
\end{figure*}


\paragraph{Task Load:} Over the whole week, \textsc{vr} induced a significantly higher taskload ($m=46.48$, $sd=2.64$) compared to \textsc{physical} ($m=34.37$, $sd=1.55$).
The mean task load for each \textsc{day} and \textsc{time} in both weeks is displayed in \autoref{fig:allPlots_m_a}.
The ANOVA results are displayed in \autoref{tab:resultsTLXSUSFlow}.
Due to logging errors, one participant had missing data and was therefore excluded from the analysis. There were no interaction effects between \textsc{day} and \textsc{environment}. 
This result indicates that participants experienced a significantly higher perceived workload when working in VR than in the comparable physical setup. 



\begin{table*}[t]
    \centering 
    \caption{RM-ANOVA results for subjective measures. d$f_1$ = d$f_{effect}$ and d$f_2$ = d$f_{error}$.}
    \small
    \setlength{\tabcolsep}{5pt}
        
        \begin{tabular}{|c||c|c|c|c|c||c|c|c|c|c||c|c|c|c|c|}
            \hline 
            &\multicolumn{5}{|c|}{Task Load} &\multicolumn{5}{|c|}{System Usability} &\multicolumn{5}{|c|}{Flow} \\
            \cline{2-16} 
            & d$f_{1}$ & d$f_{2}$ & F & p &  $\eta^2_p$  
            & d$f_{1}$ & d$f_{2}$ & F & p &  $\eta^2_p$ 
            & d$f_{1}$ & d$f_{2}$ & F & p &  $\eta^2_p$ \\ 
            \hline 
            Environment & $1$   & $14$  & $12.03$  &    \cellcolor{lightgray}$.003$ &  $.46$    
            & $1$   & $15$ & $21.14$ & \cellcolor{lightgray}$<.001$ &  $.58$    
            & $1$   & $12$  & $7.72$ & $.02$ &  $.39$   \\ 
            \hline 
            Environment*Day &  $4$  & $56$  & $2.04$  & $.10$   & $.13$     
            & $4$  & $60$  & $.75$  & $.57$   & $.05$    
            & $4$  & $48$  & $1.37$  & $.26$   & $.10$     \\ 
            
            Environment*Time & $1$   & $14$  & $.08$  & $.78$   & $.01$   
            & $1$   & $15$  & $.01$  & $.91$   & $<.001$    
            & $1$   & $12$  & $.33$  & $.57$   & $.03$      \\ 
            \hline 
        \end{tabular}

        \begin{tabular}{|c||c|c|c|c|c||c|c|c|c|c||c|c|c|c|c||c|c|c|c|c|}
            \hline 
            &\multicolumn{5}{|c|}{Perceived Productivity} &\multicolumn{5}{|c|}{Frustration} &\multicolumn{5}{|c|}{Positive Affect} &\multicolumn{5}{|c|}{Negative Affect} \\
            \cline{2-21} 
            & d$f_{1}$ & d$f_{2}$ & F & p &  $\eta^2_p$  
            & d$f_{1}$ & d$f_{2}$ & F & p &  $\eta^2_p$
            & d$f_{1}$ & d$f_{2}$ & F & p &  $\eta^2_p$
            & d$f_{1}$ & d$f_{2}$ & F & p &  $\eta^2_p$ \\ 
            \hline 
            Environment & $1$ & $15$ & $1.01$ & $.01$ &  $.46$    
            & $1$ & $15$ & $11.70$ & \cellcolor{lightgray}$.003$ &  $.44$  
            & $1$   & $12$  & $2.14$  &    $.17$ &  $.15$ & 
            $1$  & $12$  & $14.44$  & \cellcolor{lightgray}$.003$  & $.55$  \\
            \hline 
            Environment*Day & $4$ & $60$ & $1.16$ & $.34$ &  $.07$     
            & $4$ & $60$ & $.19$ & $.94$ &  $.01$ 
            & $4$  & $48$  & $.83$  & $.52$   & $.06$  & 
            $4$  & $48$  & $4.11$  & \cellcolor{lightgray}$.006$  & $.25$     \\
            Environment*Time & $1$ & $15$ & $6.96$ & \cellcolor{lightgray}$.02$ &  $.32$  
            & $1$ & $15$ & $.02$ & $.88$ &  $.001$ 
            & $1$   & $12$  & $.05$  & $.82$   & $.004$  & 
            $1$   & $12$   & $3.15$   & $.10$   & $.21$      \\  
            \hline 
        \end{tabular}

    \label{tab:resultsTLXSUSFlow}
\end{table*}

\paragraph{System Usability:} Over the whole week, \textsc{physical} resulted in a significantly higher system usability ($m=73.88$, $sd=1.49$) compared to \textsc{vr} ($m=54.71$, $sd=1.32$).
The mean system usability for each \textsc{day} and \textsc{time} in both weeks is displayed in figure \autoref{fig:allPlots_m_a}. The ANOVA results are displayed in \autoref{tab:resultsTLXSUSFlow}. There were no interaction effects between \textsc{day} and \textsc{environment}. 
We can conclude from this result that participants found the VR working arrangement far less usable than the comparable physical setup.

\paragraph{Flow:} Over the whole week, \textsc{physical} resulted in a significantly higher flow  ($m=4.76$, $sd=0.15$) compared to \textsc{vr} ($m=4.11$, $sd=0.18$). The mean flow score for each \textsc{day} and \textsc{time} in both weeks is displayed in \autoref{fig:allPlots_m_a}. The ANOVA results are displayed in \autoref{tab:resultsTLXSUSFlow}. Due to logging errors, we lost the data of three participants for this measure. There were no interaction effects between \textsc{day} and \textsc{environment}. 
This result suggests that working in VR did not support participants' focus and sense of active engagement in their work activity in a better way compared to \textsc{physical}.


\paragraph{Perceived Productivity:} Over the whole week, \textsc{physical} induced a higher level of perceived productivity ($m=4.89$, $sd=0.23$) compared to \textsc{vr} ($m=4.11$, $sd=0.28$). The mean productivity scores for each \textsc{day} and \textsc{time} in both weeks are displayed in \autoref{fig:allPlots_m_a}. The ANOVA results are displayed in \autoref{tab:resultsTLXSUSFlow}. In addition, an interaction effect between \textsc{time} and \textsc{environment} was detected, but 
post hoc tests, comparing \textsc{vr} with \textsc{physical} indicated that \textsc{vr} resulted in lower perceived productivity for both times (\textsc{morning}, \textsc{afternoon}).
This result suggests that working in \textsc{vr} did lead to a significant decrease in perceived productivity compared to \textsc{physical}.


\paragraph{Frustration:} Over the whole week, \textsc{vr} resulted in a significantly higher score for frustration ($m=3.49$, $sd=0.34$) compared to \textsc{physical} ($m=2.45$, $sd=0.26$). The mean frustration scores for each \textsc{day} and \textsc{time} in both weeks are displayed in \autoref{fig:allPlots_m_a}. The ANOVA results are displayed in \autoref{tab:resultsTLXSUSFlow}. There were no interaction effects between \textsc{day} and \textsc{environment}. This result suggests that working in \textsc{vr} did lead to a significant increase in frustration compared to \textsc{physical}.  

\paragraph{Presence:} Questions about presence only make sense in the \textsc{vr} condition. Therefore, this measure is merely descriptive. The mean presence scores for each \textsc{day} and \textsc{time} are displayed in \autoref{fig:allPlots_m_a}. The presence score ranges from 0 to 6, with 6 being the maximal amount of presence perceived by participants. Among all participants over the whole week the mean total presence score was $3.06~ (sd=1.15)$.
The sub-scores for spatial presence was $3.66~(sd=1.4)$, for involvement $2.39~(sd=1.01)$ and for experienced realism $2.49~(sd=1.21)$.




\paragraph{Positive/Negative Affect:} Over the whole week, \textsc{vr} resulted in a significantly higher negative affect ($m=11.97$, $sd=1.04$) compared to \textsc{physical} ($m=11.11$, $sd=0.52$). No such effect could be detected for positive affect. The analysis also indicated a significant interaction effect between \textsc{environment} and \textsc{day} on negative affect, which was not confirmed in post hoc tests. The mean scores for positive and negative affect for each \textsc{day} and \textsc{time} for both weeks are displayed in \autoref{fig:allPlots_m_a}. Due to data collection errors 4 data points of 3 different participants are missing, so these participants had to be removed to conduct the ANOVA. The ANOVA results are displayed in \autoref{tab:resultsTLXSUSFlow}.
There were no significant differences in the trendline slopes for negative affect over days ($p=0.082$).
These findings indicate that working in \textsc{vr} was more detrimental to participants' moods than working in the physical setup.

\begin{figure*}[t]
	\centering 
	\includegraphics[width=2.0\columnwidth]{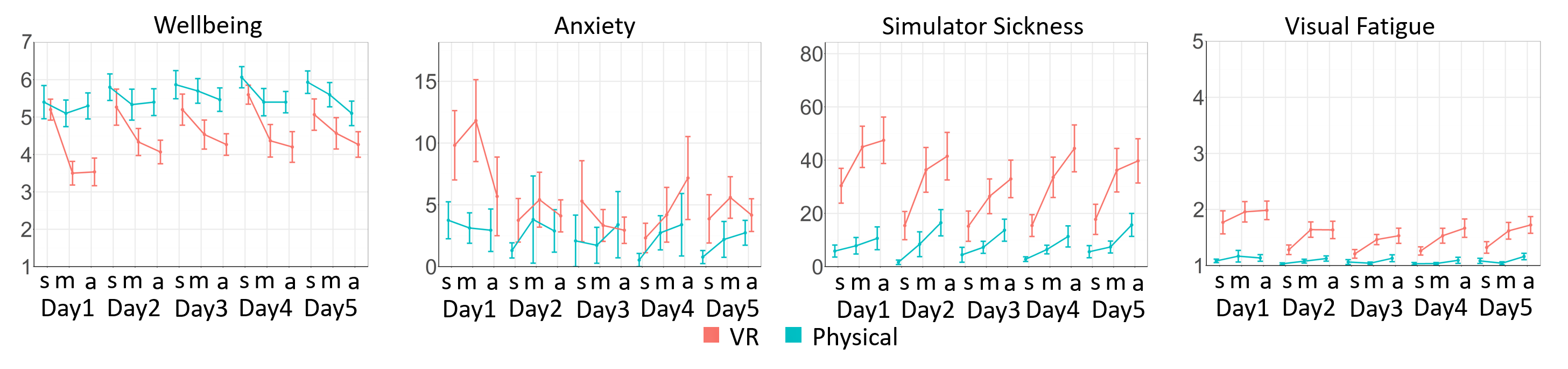}
	\caption{Average values and standard error for subjective measures at the start of the day (s), in the morning (m) and afternoon (a).}
	\label{fig:allPlots_sma}
\end{figure*}

\paragraph{Wellbeing:} Over the whole week, \textsc{physical} resulted in a significantly higher wellbeing ($m=5.31$, $sd=0.34$) compared to \textsc{vr} ($m=4.25$, $sd=0.59$). The mean scores for wellbeing for each \textsc{day} and \textsc{time} in both weeks are displayed in \autoref{fig:allPlots_sma}. The ANOVA results are displayed in \autoref{tab:resultsSTAISSQVF}. Due to logging errors, data of one participant is missing. There were no interaction effects between \textsc{day} and \textsc{environment}. However, an interaction effect between the \textsc{environment} and \textsc{time} was detected. Post hoc tests, comparing the \textsc{vr} to \textsc{physical} condition for each of the three times (\textsc{start}, \textsc{morning}, \textsc{afternoon}) indicated that the \textsc{vr} condition results in a significantly lower wellbeing in the \textsc{morning} ($V=10$, $p=0.015$, $r=0.71$, $mean_{VR}=4.07$, $sd_{VR}=1.4$, $mean_{Physical}=5.26$, $sd_{Physical}=1.42$) and \textsc{afternoon} ($V=11$, $p=0.01$, $r=0.74$, $mean_{VR}=3.91$, $sd_{VR}=1.2$, $mean_{Physical}=5.15$, $sd_{Physical}=1.27$), but not at \textsc{start}. 


\paragraph{Anxiety:} Over the whole week, \textsc{vr} resulted in a significantly higher anxiety ($m=5.3$, $sd=5.84$) compared to  \textsc{physical} ($m=2.42$, $sd=5.34$). The mean scores for anxiety for each \textsc{day} and \textsc{time} in both weeks are displayed in \autoref{fig:allPlots_sma}. The ANOVA results are displayed in \autoref{tab:resultsSTAISSQVF}. There were also interaction effects between \textsc{environment} and both \textsc{day} and \textsc{time}, which were not confirmed in post-hoc tests. 
Also, there were no significant differences in the trendline slopes over days ($p=0.21$).
The findings suggest that working in VR elevated participants' feelings of anxiety.

\begin{table*}[t]
    \centering 
    \caption{RM-ANOVA results for Wellbeing, Anxiety, Simulator Sickness and Visual Fatigue. d$f_1$ = d$f_{effect}$ and d$f_2$ = d$f_{error}$.}
    \small
    \setlength{\tabcolsep}{5pt}
        
        \begin{tabular}{|>{\centering}m{1.8cm}||>{\centering}m{0.2cm}|c|c|c|>{\centering}m{0.25cm}||>{\centering}m{0.2cm}|c|c|c|>{\centering}m{0.25cm}||>{\centering}m{0.2cm}|c|c|c|>{\centering}m{0.25cm}||>{\centering}m{0.2cm}|c|c|c|c|}
            \hline 
            &\multicolumn{5}{|c|}{Wellbeing}&\multicolumn{5}{|c|}{Anxiety} &\multicolumn{5}{|c|}{Simulator Sickness} &\multicolumn{5}{|c|}{Visual Fatigue}  \\
            \cline{2-21} 
            & d$f_{1}$ & d$f_{2}$ & F & p &  $\eta^2_p$
            & d$f_{1}$ & d$f_{2}$ & F & p &  $\eta^2_p$
            & d$f_{1}$ & d$f_{2}$ & F & p &  $\eta^2_p$
            & d$f_{1}$ & d$f_{2}$ & F & p &  $\eta^2_p$\\ 
            \hline 

            Environment &  $1$ & $14$ & $13.34$ & \cellcolor{lightgray}$.002$ &  $.49$ 
            & $1$   & $15$  & $20.35$  &    \cellcolor{lightgray}$<.001$ &  $.58$ 
            & $1$   & $15$  & $24.34$  &    \cellcolor{lightgray}$<.001$ &  $.62$  
            & $1$   & $15$  & $26.30$  &    \cellcolor{lightgray}$<.001$ &  $.64$  \\ 

            \hline 
            
            Environment*Day & $4$ & $56$ & $.70$ & $.59$ &  $.05$ 
            &  $4$  & $60$  & $5.98$  &  \cellcolor{lightgray}$<.001$   & $.28$   
            &  $4$  & $60$  & $10.32$  &  \cellcolor{lightgray}$<.001$   & $.41$  
            &  $4$  & $60$  & $12.98$  &  \cellcolor{lightgray}$<.001$   & $.46$  \\  

            Environment*Time & $2$ & $28$ & $5.70$ & \cellcolor{lightgray}$.008$ &  $.29$
            & $2$   & $30$  & $4.07$  &   \cellcolor{lightgray}$.03$   & $.21$   
            & $2$   & $30$  & $19.06$  &  \cellcolor{lightgray}$<.001$   & $.56$  
            & $2$   & $30$  & $27.10$  &  \cellcolor{lightgray}$<.001$   & $.64$  \\ 
            \hline
            
        \end{tabular}

    \label{tab:resultsSTAISSQVF}
\end{table*}

\paragraph{Simulator Sickness:} Over the whole week, simulator sickness scores were significantly higher in  \textsc{vr}  ($m=34.3$, $sd=10.16$) compared to  \textsc{physical}  ($m=9.21$, $sd=4.47$). According to Stanney et al. \cite{stanney1997cybersickness} these symptoms in \textsc{vr} can be considered bad. The mean scores for each \textsc{day} and \textsc{time} for both weeks are displayed in \autoref{fig:allPlots_sma}.
The ANOVA results are displayed in \autoref{tab:resultsSTAISSQVF}.
Interaction effects between the \textsc{environment} and both \textsc{day} and \textsc{time} were detected. Post hoc tests revealed a significant difference between \textsc{vr} and  \textsc{physical} condition on all days
and for all three time periods. 
Also, there were no significant differences in the trendline slopes over days ($p=0.078$).
These findings suggest that \textsc{vr} leads to substantial simulator sickness symptoms over the course of the week.


\paragraph{Visual Fatigue:} Over the whole week, \textsc{vr} ($m=1.61$, $sd=0.22$) resulted in a significantly higher visual fatigue than \textsc{physical} ($m=1.09$, $sd=0.05$). The mean visual fatigue scores for each \textsc{day} and \textsc{time} in both weeks are displayed in \autoref{fig:allPlots_sma}. The ANOVA results are displayed in \autoref{tab:resultsSTAISSQVF}. Additionally, interaction effects between all variables were found.
Post hoc tests revealed a significant difference between \textsc{vr} and \textsc{physical} on every day 
and for all time periods (\textsc{start}, \textsc{morning} and \textsc{afternoon}).
Comparing the slopes of lines fitted through the mean ratings for each day, we found that in \textsc{vr} ($m=-0.069$, $sd=0.13$) visual fatigue decreased at a significantly higher rate than in \textsc{physical} ($m=-0.01$, $sd=0.03$) during the course of the week ($p=0.04$, $Cohen's~d= -0.47$). This suggests that while visual fatigue was substantially higher in \textsc{vr}, it also decreased significantly faster that \textsc{physical}. 


\begin{figure*}[t]
	\centering 
	\includegraphics[width=2\columnwidth]{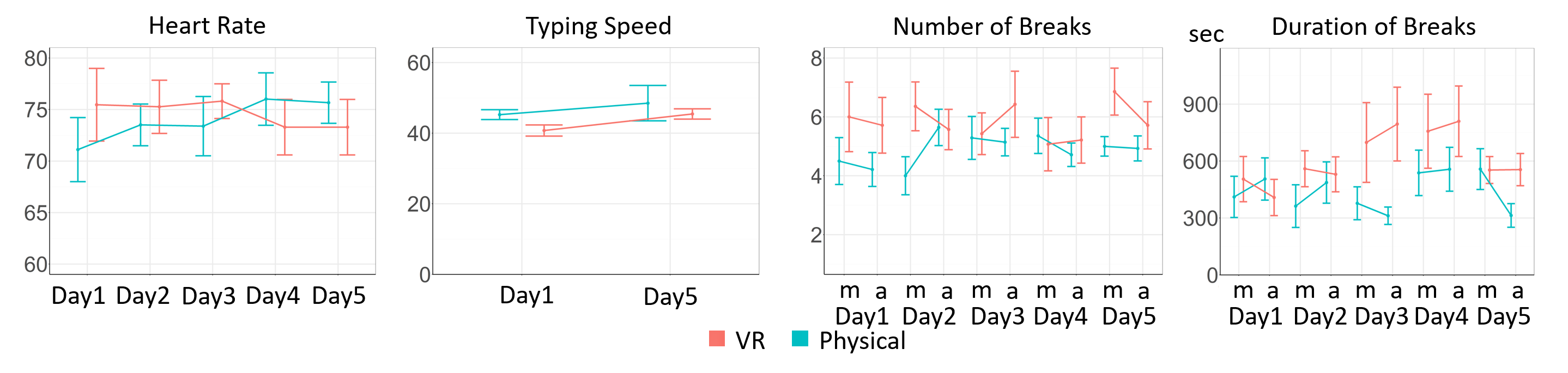}
	\caption{Average values and standard error for objective measures in the morning (m) and afternoon (a).}
	\label{fig:all_cont_plots}
\end{figure*}



\paragraph{Heart Rate:} Statistical tests showed no significant influence of \textsc{environment} on heart rate. The mean heart rate among all participants for each \textsc{day} and \textsc{time} for both weeks are displayed in \autoref{fig:all_cont_plots}. The ANOVA results are displayed in \autoref{tab:resultsBreakHartType}.  There was a significant interaction effect between \textsc{day} and \textsc{environment}. It can be seen in \autoref{fig:all_cont_plots} that on \textsc{day4} and \textsc{day5} the average heart rate is higher in \textsc{physical} condition while it is higher in \textsc{vr} for the first three days. Post hoc tests, however, could not identify significant differences. Due to logging errors we lost the data of 6 participants.






\paragraph{Break Times:} No significant effect of \textsc{environment} could be detected on the number of breaks. However, the average duration of a break was significantly higher in \textsc{vr} ($m=617.05s$, $sd=137.42$) compared to \textsc{physical} ($m=442.121s$, $sd=98.1$). The mean number of breaks and the average duration are displayed in \autoref{fig:all_cont_plots}. The ANOVA results are displayed in \autoref{tab:resultsBreakHartType}.

For acquiring break times, we used the videos and head-/marker-tracking algorithms, 
and considered it a break if no head/marker could be detected.
We were interested only in actual breaks that participants used to rest. Hence, in this analysis, we only consider a break 
if in \textsc{vr} the participant took off the headset for more than 30 seconds and in \textsc{physical} if the participant turned away from the screen for more than 30 seconds. We do not consider shorter breaks, since the videos indicated, that these are mainly due to participants quickly turning around, adjusting the headset, picking something up or behaving in a way that results in tracking being lost. For all break times generated by the tracking algorithms, we manually verified them.
Additionally, we compared the resulting break times with the ManicTime data, which logged all time frames where the user was inactive for more than 10 minutes, so we do not miss any major breaks, which might have not been detected by the tracking algorithms (e.g., a user facing the camera while using a smartphone instead of the work PC).



\begin{table*}[t]
    \centering 
    \caption{RM-ANOVA results for breaks. d$f_1$ = d$f_{effect}$ and d$f_2$ = d$f_{error}$.}
    \small
    \setlength{\tabcolsep}{5pt}
        
        \begin{tabular}{|c||c|c|c|c|c||c|c|c|c|c||c|c|c|c|c||c|c|c|c|c|}
            \hline 
            &\multicolumn{5}{|c|}{Number of Breaks} &\multicolumn{5}{|c|}{Break Duration} &\multicolumn{5}{|c|}{Heart Rate} &\multicolumn{5}{|c|}{Typing Speed}  \\
            \cline{2-21} 
            & d$f_{1}$ & d$f_{2}$ & F & p &  $\eta^2_p$  
            & d$f_{1}$ & d$f_{2}$ & F & p &  $\eta^2_p$
            & d$f_{1}$ & d$f_{2}$ & F & p &  $\eta^2_p$
            & d$f_{1}$ & d$f_{2}$ & F & p &  $\eta^2_p$
             \\ 
            \hline 

            Environment & $1$   & $13$  & $0.87$  &    $0.37$ &  $0.06$  
            & $1$ & $13$ & $12.27$ & \cellcolor{lightgray}$.004$ &  $.49$
            & $1$   & $9$  & $.08$  &    $.78$ &  $.009$
            & $1$ & $15$  & $8.46$  & \cellcolor{lightgray}$.01$  & $.36$
             \\ 

            \hline 

            Environment*Day &  $4$  & $52$  & $1.57$  & $0.20$   & $0.11$   
            & $4$ & $52$ & $2.01$ & $0.11$ &  $0.13$
            & $4$  & $36$  & $1.77$  & $.16$   & $.17$
            & $1$  & $15$   & $.31$   & $.59$   & $.02$
             \\ 
            \cline{12-21}

            Environment*Time & $1$   & $13$  & $0.71$  & $0.41$   & $0.05$  
            & $1$ & $13$ &  $<0.001$ & $0.98$ &  $.0$ 
            &   \multicolumn{10}{|c}{ }     
             \\ 
            
            \cline{1-11}

        \end{tabular}

    \label{tab:resultsBreakHartType}
\end{table*}

\paragraph{Typing Speed:} Both \textsc{environment} and \textsc{day} had an influence on typing speed such that \textsc{physical} condition resulted in a significantly faster typing speed ($m=46.88$, $sd=20.92$) than \textsc{vr} ($m=43.09$, $sd=23.98$). On \textsc{day5} ($m=46.97$, $sd=21.42$) participants were typing significantly faster than on \textsc{day1} ($m=43.0$, $sd=23.52$) in both conditions.
The RM-ANOVA results are displayed in \autoref{tab:resultsBreakHartType} and the means among all participants are displayed in \autoref{fig:all_cont_plots}. There were no interaction effects between \textsc{day} and \textsc{environment}. These results are in line with prior work suggesting a mild performance drop when typing with physical keyboards in VR \cite{grubert2018text}.

\paragraph{Interviews:} At the end of each week, we gained additional information from the participants through an interview in which we asked them about how they felt during the week, what they liked or disliked, which problems occurred, and what they would improve. At the end of the second week, they were also asked which \textsc{environment} they preferred and if they could imagine using VR for work in the future.




For the \textsc{vr} condition, 11 participants disliked the comfort or proposed to increase it  
(P01-P03, P05, P07-P11, P13, P15). Major issues mentioned include the weight of the HMD and its pressure against the face. 
P03 and P05 mentioned that the peripheral view was not satisfactory, so they had to move their heads more often. 
P03 and P05 also pointed out that they needed to take off their headsets for drinking or eating because they were afraid to spill something, and P04 said that such tasks were harder with the HMD on. P05 and P13 missed the ability to write something down on paper. Participants also mentioned technical details that could be improved. For example, removing the headset sometimes made it necessary to reset the position of the virtual screen (using a menu in the Oculus system software) or the remote connection (P03, P04, P13). 
In addition, hand gestures were sometimes falsely recognized while typing (P06, P02, P12), resulting in an involuntary selection action. Also, four participants mentioned the tracking of the keyboard could be improved (P06, P09, P13, P11).

Three participants mentioned that the study was too long (P07, P09, P10). Seven participant said they felt tired during \textsc{vr} condition (P07, P09, P10, P13 - P16) and only two (P09, P13) during \textsc{physical} condition. However, the main reason for this seemed to be deviation from their normal schedules. For example, P13 said ``Maybe I was more tired than usual, because I usually do not work for that much time continuously''. Five participants (P04, P06, P12, P13, P16) mentioned that they got used to wearing the HMD during the \textsc{vr} condition; however, P06 and P09 also mentioned that the second half of the day was usually harder. P13 mentioned ``I did more work than I usually do,'' while P12 felt that he ``was not as productive, because of the low resolution and keyboard''. P03 revealed that he ``had a blurry vision when driving home on the first day''. 
Seven participants (P03, P04, P06 - P08, P13, P16) mentioned that they felt ``alright'' during the \textsc{physical} condition, while only P08 explicitly stated that she felt alright in the \textsc{vr} condition.

Nine participants (P01, P03, P05, P06, P09, P12 - P15) liked that the isolation in the \textsc{vr} condition allowed them to concentrate more on the tasks at hand, because they were not distracted, especially in combination with music from their private headphones. However, this could also have drawbacks, and as P01, P06 and P08 mentioned, the \textsc{vr} condition was ``a bit scary,'' because they could not see the presence of other people in the real world (see also \cite{mcgill2015dose}). 


P12 also mentioned that ``without [private] music turned on, I was trying to guess what was happening around me''.
P01 and P13 said that they even forgot that they were wearing HMDs when concentrating hard on their work and P11 mentioned that the experience and movement in VR felt natural. Four participants (P04, P07, P08, P12) specifically mentioned that they liked to try out and experience VR in a work context.
P09 liked the privacy that VR offered, as ``nobody can see what you are doing''. While P10 and P13 liked to relax in the virtual environment, P13 also ``liked to look around when taking a break and just looking at empty space''.
On the other hand, P06, P08 and P12 mentioned that they felt more comfortable seeing the real surrounding and P09 liked to look somewhere else and not on a display when resting.

Only three participants (P01, P11, P13) preferred the \textsc{vr} condition. P01 felt more relaxed in VR, P13 liked the isolation and was able to do more work, while P11 liked it because he already knew the system/study from the \textsc{physical} condition week.
P02, P05 and P12 preferred the \textsc{physical} condition because it was the same as the \textsc{vr}, but without downsides such as the heavy HMD. 
Others preferred the \textsc{physical} condition because it was more familiar (P04, P08, P09), it felt less limiting (P03), it did not require wearing the HMD (P07), it was easier (P10, P16) and it allowed to focus more on work as she ``did not have to tackle problems with the system'' (P15). Nevertheless, P06 and P05 added that \textsc{vr} was more exciting.

All participants could imagine using VR for work in the future if some conditions are met, such as having lighter HMDs with higher resolution and being able to have multiple displays. Also, all participants mentioned that they could imagine using VR for a limited amount of time (on train rides or for certain tasks).
Regarding time, P12 mentioned that ``in VR I had 45 minutes of high performance and then 3 hours of headache''. P06 suggested that using VR could improve ergonomics, because displays are adjustable, as well as that VR would be good for working at home to separate work from personal life. P03 and P07 mentioned that they prefer not to sit in one place all day and they usually like to walk around and talk to coworkers. P04 would also like to have the possibility to play games with other users in VR during breaks and to have their phone integrated in VR. P16 mentioned that the isolation in VR could hinder collaboration with colleagues.
For both conditions, participants mentioned concerns about the keyboard, touchpad, screen resolution and delay induced by the remote desktop. Please note that this is expected and a result of making the two interfaces comparable with available hardware.

A week after the experiment, participants were also asked if they observed any effects after completing the \textsc{vr} week. P04 mentioned that sometimes during the weekend, she felt as if she was still wearing the HMD. P16 and P15 still had a feeling of dry eyes after finishing the use of VR and P15 felt sleepy and dizzy for about 2 hours afterwards.  Also, P01 mentioned she felt a swelling of the face around the eyes and her neck and shoulders were stiff. P16 and P06 were ``amazed by how detailed the real world is after removing HMD''. P02 and P06 felt their skin suffered after wearing the headset for one week. All other participants did not report major effects.

\subsection{Dropout}
In total, two participants decided to drop out. The first participant who dropped out experienced regular migraines and mentioned that the weight of the headset triggered them. Therefore, this participant dropped out after two blocks of work (four hours). The second participant who  dropped out explained that due to the weight of the headset it was not possible to sit in a relaxed position. In addition, this participant mentioned that not having access to the usual setup reduced motivation and productivity. After approximately two hours this participant experienced anxiety and felt nauseated and disoriented, resulting in the participant dropping out from the study. 

\section{Discussion}




In the presented study, we have examined the experience of users working for one work week in VR and a comparable physical setup. 
To control for various factors, such as screen size, input device and working conditions, our experimental protocol enforced as similar configurations in both conditions as possible, while considering work experiences that are accessible to a wide number of users using today's commercial off-the-shelf VR solutions (an Oculus Quest2 with accompanying Logitech keyboard). Given the limitations of current technology and the fact that VR provides a virtual approximation of the real environment, we did not expect the \textsc{vr} condition to outperform the \textsc{physical} condition 
which is also confirmed by the results.
However, the quantified results of the studied VR experience that is comparable to a physical one, can serve as a baseline for future optimized VR systems. In fact, the extent to which some of our measures diverged between \textsc{vr} and \textsc{physical} is notable. For example, \textsc{vr} clearly resulted in below average system usability scale ratings, while \textsc{physical} resulted in above average ratings, even though both systems used the same input devices and had comparable screen real estates for carrying out work. Similarly, while it is expected that a VR system induces higher simulator sickness ratings than a non-VR system, the absolute values of the SSQ ratings indicate that \textsc{vr} corresponded to the worst category of simulator sickness \cite{stanney1997cybersickness}\footnote{Please note that this categorization is based mostly on military simulators and that researchers discuss about the associated challenges \cite{bimberg2020usage}.}. We find this surprising given that we utilized a setup (Oculus Quest2, Logitech Keyboard), which can be considered widespread among consumers and professionals alike. These high ratings of simulator sickness could be observed throughout the week, even though we note that the SSQ ratings decreased slightly over the week. On the other hand, not all differences are as concerning. While the self-rated task load in \textsc{vr} was approximately 35\% higher compared to \textsc{physical} condition, it is still within the 50th percentile of ratings of computer activities \cite{grier2015high}.

\hl{We also examined, if there are any differences between touch-typists and non-touch-typists. We divided the participants into two groups based on their need to look at the keyboard while typing which they reported in the demographic questionnaire on a seven-point likert scale. Six participants answered either 1 (never) or 2 and were therefore considered touch-typists. An analysis with \textsc{touchtypist} as a between-subjects factor revealed no significant main effects of this variable. This suggests that the need to look at the keyboard while typing does not have a significantly negative influence on the results.}


When fitting a linear model to the data, we observed that any improvements across most measures in both conditions are not significant. Still, examining the development of the scores over the week can serve as an indication of a possible emerging trend in the future.
When inspecting the graphs, it is possible to observe a rapid adaptation of users to the \textsc{vr} condition. Within a day or two, many of the scores for \textsc{vr} improved. 
At this stage, we do not know if this improvement was a result of participants' brains adapting to the new condition or if they overcame the initial expectations that people had previously about VR.
Another effect observable in the results was a gradual accumulation of some exhaustion across the week. We observed this effect in some of the measures of \textsc{vr} (specifically, regarding task load, simulator sickness). We also observed such an effect in the \textsc{physical} condition, although with slower growth, which may hint that this factor might be independent of the environment and instead more related to the duration of the experiment.



It is clear that there is still a long way to go for the development of more comfortable hardware. There are already HMDs that offer higher resolution displays, faster refresh rates, variable focal distance displays, and wider field of view. We anticipate future HMDs will be available in a form factor similar to conventional glasses, and be lighter and allow the flow of air around users' face. We expect that such hardware will further reduce the gap between \textsc{vr} and the \textsc{physical} conditions. Some of the more mundane issues encountered by the participants in the \textsc{vr} condition are relatively straightforward to address. For example, multiple participants complained about the keyboard periodically vanishing from the VR environment, the relative position of the home environment shifting, and hand movements on the keyboard being inadvertently recognized as input gestures. To address this, a dedicated ‘work’ mode may be appropriate, allowing device tracking and gesture recognition subsystems to operate in a more persistent manner when the user is known to be engaged in seated work.

The comments from participants in the interviews highlight the challenge of implementing an effective and enjoyable VR working experience given the potential influence of personal preference.
We note that several participants appreciated how working in VR helped isolate them from their physical workspace and enabled periods of greater focus, separation and privacy.
Conversely, other participants had negative experiences due to this isolation, whether due to a feeling of unease produced by an inability to perceive who is nearby or the obstacles the setup presents for face-to-face collaboration.
Nevertheless, despite the generally negative experience reported by the majority of participants, all commented that they could imagine using VR for at least some work tasks or at least a portion of the day. This hints to the future when knowledge workers will combine two modes depending on the needs for the work at hand. 

\subsection{Limitations and Future Work}
Conducting a complex in-situ study carried out across an entire working week is inherently intertwined with variables that are outside of our control. Therefore, many of the measures depend on factors that we cannot fully control. For example, reflections on frustration, perceived productivity or ability to concentrate may be influenced by the type of work being performed.
There are also a number of other aspects of the study that should be considered when interpreting the findings. 
First, we discovered that although the overall task load was higher in \textsc{vr}, task load was at a relatively high level in both conditions (but still within the bound of comparable computer work~\cite{stanney1997cybersickness}). This is likely a consequence of the limitations of the setup common to both conditions, such as no use of a mouse, and a display set at a relatively low resolution. Second, many of our measures are based on participants' subjective responses. However, as observed by Wille et al.~\cite{wille2014prolonged}, there can be a disconnect between objective physiological effects and subjective user ratings. In terms of eye strain, Wille et al.\cite{wille2014prolonged} suggest that this disconnect may be influenced by the level of familiarity a user has with the technology. If true, this may serve to explain some of the reduction seen in some measures over the first few days of the study. Third, our understanding of other factors related to working in VR, and how they impact the user experience, is still emerging. Shen et al.~\cite{shen2019mental} suggest that VR allows for the use of more attention resources. This may help explain the experience of P12, who commented that he experienced a high efficiency for 45 minutes and then a headache for three hours. If VR does indeed allow for the use of more attention resources, steps should be taken to avoid overloading users, particularly when they are still acclimating to working in VR. Finally, we only presented selected analyses of the collected data. As we release the collected anonymized data, future work could investigate the data further. Since we employed a widely available commercially available hardware and software solution, we would also hope that further researchers could add to the data set by replicating the study.

This paper has studied the effects of working in VR compared to a regular working environment at one particular operating point with experimental parameters set to align as much as possible between both environments, given the constrains of using a commodity off-the-shelf VR system. We hope this work will stimulate further work at different operating points, investigating how some of the quantified effects we observed in this study may possibly change if the VR condition is allowed to deviate from the operating point of a regular working environment, for instance, by providing flexible solutions to allow VR users making maximum use of the available VR space and novel VR interaction techniques to make VR interaction more comfortable.

\hl{Additionally, the duration of the study, the need to exercise control over the work environment, and the fact that participants were required to perform their standard work tasks largely restricted feasible recruitment to individuals already embedded within the three different university sites. Such individuals may inadvertently be more forgiving of the deficiencies of the setup. Future work is required to look at how a broader population may experience working in VR.}

We also see interesting further work in examining stress and heart rate more closely when working in different VR environments. There are also open issues around the social acceptability of working in VR for a prolonged amount of time, as well as users possibly feeling isolated or having difficulties in collaborating with their colleagues.




    
    
    
    

\section{Conclusions}
In this paper, we have studied the effects of working in VR for an entire workweek. While VR has been repeatedly pitched as providing new exciting possibilities for modern knowledge work, in practice the potential advantages of virtual work environments can only be used if it is feasible to work in a virtual environment for an extended period of time. Prior to this work, there were only limited studies of long-term effects of working in VR. We reported the results from a comparative study with 16 participants working for an entire workweek in both VR and in a baseline physical desktop environment. As a first study of this kind and scale, we deliberately opted to design these conditions to be as similar as possible to allow as many quantitative comparisons as possible. Therefore, the study did not present the participants with the best possible VR system but instead a setup that delivered a comparable experience to working in a physical desktop environment. The study revealed that, as expected, VR resulted in significantly worse ratings across most measures. For example, VR resulted in below average system usability scale ratings while the physical environment resulted in above average ratings. We also found that VR resulted in the worst category of simulator sickness although the severity decreased slightly across the week. \hl{However two participants even dropped out on the first \textsc{vr} day, due to migraine, nausea and anxiety.} Nevertheless, there was some indication that participants gradually overcame negative first impressions and initial discomfort. Overall, this study helps laying the groundwork for subsequent research, highlighting current shortcomings and identifying opportunities for improving the experience of working in VR. We hope this work will stimulate further research investigating longer-term productive work in-situ in VR.



\balance


\bibliographystyle{abbrv-doi}

\typeout{}

\balance
\bibliography{template}
\end{document}




\maketitle


\begin{figure}[t]
	\centering 
	\includegraphics[width=1\columnwidth]{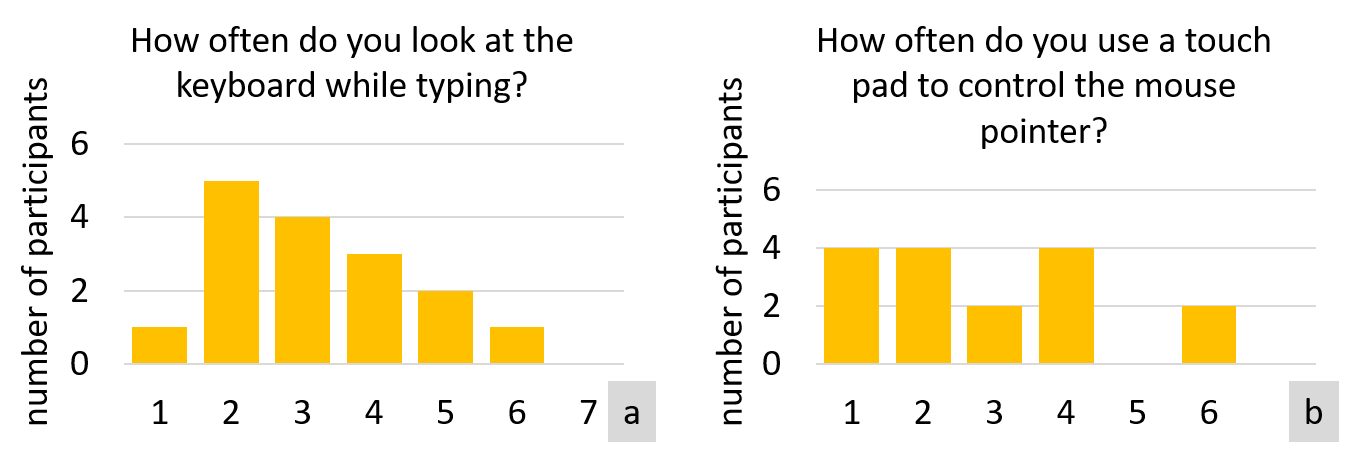}
	\caption{How often participants usually look at keyboard while typing (a) and how often they use a touch pad to control the mouse pointer (b). Scale from 1 (never) to 7 (all the time).}
	\label{fig:howOften}
\end{figure}

\begin{figure*}[!htbp]
	\centering 
	\includegraphics[width=2\columnwidth]{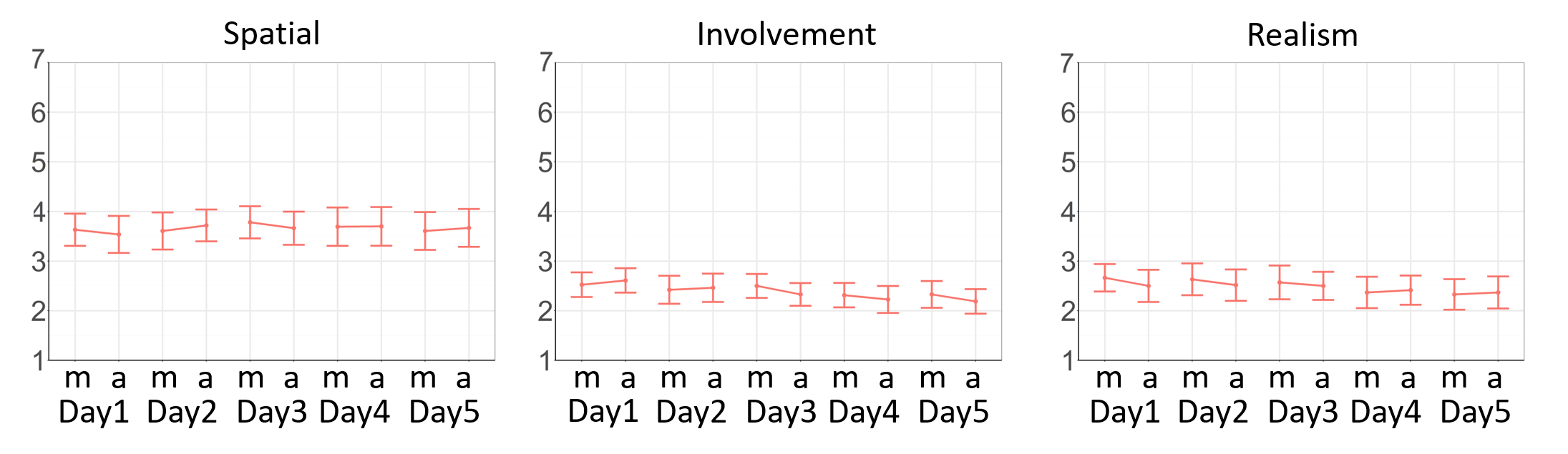}
	\caption{Subscales of Presence Questionnaire (IPQ).}
	\label{fig:howOften}
\end{figure*}

\begin{table*}[t]
    \centering 
    \tiny
    \setlength{\tabcolsep}{5pt}
        \caption{RM-ANOVA results for Task Load, system Usability, Flow, Productivity, Frustration, Positive and Nnegative Affect. d$f_1$ = d$f_{effect}$ and d$f_2$ = d$f_{error}$.}
        \begin{tabular}{|c||c|c|c|c|c||c|c|c|c|c||c|c|c|c|c|}
            \hline 
            &\multicolumn{5}{|c|}{Task Load} &\multicolumn{5}{|c|}{System Usability} &\multicolumn{5}{|c|}{Flow} \\
            \hline 
            & d$f_{1}$ & d$f_{2}$ & F & p &  $\eta^2_p$  & d$f_{1}$ & d$f_{2}$ & F & p &  $\eta^2_p$  & d$f_{1}$ & d$f_{2}$ & F & p &  $\eta^2_p$ \\ 
            \hline 
            
            Environment & $1$   & $14$  & $12.03$  &    \cellcolor{lightgray}$0.003$ &  $0.46$    
            & $1$   & $15$ & $21.14$ & \cellcolor{lightgray}$<0.001$ &  $0.58$    
            & $1$   & $12$  & $7.72$ & $0.02$ &  $0.39$   \\ 
            
            Day & $4$ & $56$ & $1.42$ & $0.24$ &  $0.09$  
            & $4$ & $60$ & $0.12$ & $0.97$ &  $0.01$ 
            & $4$ & $48$ & $0.71$ & $0.59$&  $0.06$ \\ 
            
            Time & $1$ & $14$ & $3.71$ & $0.07$ &  $0.21$  
            & $1$ & $15$ & $0.48$ & $0.50$ &  $0.03$  
            & $1$ & $12$ & $0.75$ & $0.40$ &  $0.06$ \\ 
            \hline 
            
            Environment*Day &  $4$  & $56$  & $2.04$  & $0.10$   & $0.13$     
            & $4$  & $60$  & $0.75$  & $0.57$   & $0.05$    
            & $4$  & $48$  & $1.37$  & $0.26$   & $0.10$     \\ 
            
            Environment*Time & $1$   & $14$  & $0.08$  & $0.78$   & $0.01$   
            & $1$   & $15$  & $0.01$  & $0.91$   & $<0.001$    
            & $1$   & $12$  & $0.33$  & $0.57$   & $0.03$      \\ 
            \hline
            
            Day*Time & $4$   & $56$  & $1.62$  & $0.18$   & $0.10$   
            & $4$   & $60$  & $0.66$  & $0.62$   & $0.04$    
            & $4$   & $48$  & $0.29$  & $0.88$   & $0.02$      \\ 
            \hline
            
            Environment*Day *Time &  $4$  & $56$  & $1.28$  & $0.29$   & $0.08$   
            & $4$  & $60$  & $0.32$  & $0.86$   & $0.02$    
            & $4$  & $48$  & $0.96$  & $0.43$   & $0.07$    \\ 
            \hline 
            
            \end{tabular}

            \begin{tabular}{|c||c|c|c|c|c||c|c|c|c|c||c|c|c|c|c||c|c|c|c|c|}
            \hline 
            &\multicolumn{5}{|c|}{Productivity} &\multicolumn{5}{|c|}{Frustration} &\multicolumn{5}{|c|}{Positive Affect} &\multicolumn{5}{|c|}{Negative Affect} \\
            \hline 
            & $f_{1}$ & d$f_{2}$ & F & p &  $\eta^2_p$  
            & d$f_{1}$ & d$f_{2}$ & F & p &  $\eta^2_p$
            & d$f_{1}$ & d$f_{2}$ & F & p &  $\eta^2_p$
            & d$f_{1}$ & d$f_{2}$ & F & p &  $\eta^2_p$ \\ 
            \hline 
            Environment & $1$ & $15$ & $1.01$ & $0.01$ &  $0.46$    
            & $1$ & $15$ & $11.70$ & \cellcolor{lightgray}$0.003$ &  $0.44$  
            & $1$   & $12$  & $2.14$  &    $0.17$ &  $0.15$ & 
            $1$  & $12$  & $14.44$  & \cellcolor{lightgray}$0.003$  & $0.55$  \\
            
            Day & $4$ & $60$ & $3.06$ & $0.23$ &  $0.17$ 
            & $4$ & $60$ & $0.92$ & $0.46$ &  $0.06$  
            & $4$  & $48$  & $3.23$  & $0.02$   & $0.21$  & 
            $4$ & $48$  & $9.70$  & \cellcolor{lightgray}$<0.001$  & $0.45$ \\ 
            
            Time & $1$ & $15$ & $0.37$ & 0.55 &  $0.02$  
            & $1$ & $15$ & $5.01$ & $0.04$ &  $0.25$  
            & $1$   & $12$  & $17.12$  & \cellcolor{lightgray}$0.001$    & $0.58$  & 
            $1$  & $12$  & $3.78$  &  $0.08$ & $0.24$ \\ 
            \hline 
            Environment*Day & $4$ & $60$ & $1.16$ & $0.34$ &  $0.07$     
            & $4$ & $60$ & $0.19$ & $0.94$ &  $0.01$ 
            & $4$  & $48$  & $0.83$  & $0.52$   & $0.06$  & 
            $4$  & $48$  & $4.11$  & \cellcolor{lightgray}$0.006$  & $0.25$     \\
            
            Environment*Time & $1$ & $15$ & $6.96$ & $0.02$ &  $0.32$  
            & $1$ & $15$ & $0.02$ & $0.88$ &  $0.001$ 
            & $1$   & $12$  & $0.05$  & $0.82$   & $0.004$  & 
            $1$   & $12$   & $3.15$   & $0.10$   & $0.21$      \\ 
            
            Day*Time & $4$ & $60$ & $0.54$ & $0.70$ &  $0.04$  
            & $4$ & $60$ & $0.47$ & $0.76$ &  $0.03$ 
            & $4$    & $48$   & $2.69$  & $0.04$    & $0.18$  & 
            $4$  & $48$  & $0.27$  & $0.89$  & $0.02$      \\ 
            \hline
            Environment*Day *Time & $4$ & $60$ & $0.21$ & $0.93$ &  $0.01$   
            & $4$ & $60$ & $1.71$ & $0.16$ &  $0.10$ 
            & $4$  & $48$  & $2.88$  & $0.03$   & $0.19$   & 
            $4$  & $48$  & $0.57$  & $0.69$   & $0.05$     \\ 
            \hline 
        \end{tabular}

    \label{tab:resultsTLXSUSFlow}
    \vspace{-0.3cm}
\end{table*}

\begin{table*}[t]
    \centering 
    \tiny
    \setlength{\tabcolsep}{5pt}
        \caption{RM-ANOVA results for Wellbeing, Anxiety, Simulator Sickness and Visual Fatigue. d$f_1$ = d$f_{effect}$ and d$f_2$ = d$f_{error}$.}
        \begin{tabular}{|c||c|c|c|c|c||c|c|c|c|c||c|c|c|c|c||c|c|c|c|c|}
            \hline 
            &\multicolumn{5}{|c|}{Wellbeing}&\multicolumn{5}{|c|}{Anxiety} &\multicolumn{5}{|c|}{Simulator Sickness} &\multicolumn{5}{|c|}{Visual Fatigue}  \\
            \hline 
            & d$f_{1}$ & d$f_{2}$ & F & p &  $\eta^2_p$
            & d$f_{1}$ & d$f_{2}$ & F & p &  $\eta^2_p$
            & d$f_{1}$ & d$f_{2}$ & F & p &  $\eta^2_p$
            & d$f_{1}$ & d$f_{2}$ & F & p &  $\eta^2_p$\\ 
            \hline 

            Environment &  $1$ & $14$ & $13.34$ & \cellcolor{lightgray}$0.002$ &  $0.49$ 
            & $1$   & $15$  & $20.35$  &    \cellcolor{lightgray}$<0.001$ &  $0.58$ 
            & $1$   & $15$  & $24.34$  &    \cellcolor{lightgray}$<0.001$ &  $0.62$  
            & $1$   & $15$  & $26.30$  &    \cellcolor{lightgray}$<0.001$ &  $0.64$  \\ 

            \hline 
            
            Day & $4$ & $56$ & $4.62$ & \cellcolor{lightgray}$0.002$ &  $0.25$ 
            & $4$ & $60$ & $11.15$ & \cellcolor{lightgray}$<0.001$ &  $0.43$ 
            & $4$ & $60$ & $8.32$ & \cellcolor{lightgray}$<0.001$ &  $0.36$ 
            & $4$ & $60$ & $16.24$ & \cellcolor{lightgray}$<0.001$ &  $0.52$  \\ 
            \hline 
            
            Time & $2$ & $28$ & $38.82$ & \cellcolor{lightgray}$<0.001$ &  $0.73$ 
            & $2$ & $30$ & $3.52$ & \cellcolor{lightgray}$0.04$ &   $0.19$ 
            & $2$ & $30$ & $38.48$ & \cellcolor{lightgray}$<0.001$ &  $0.72$ 
            & $2$ & $30$ & $24.41$ & \cellcolor{lightgray}$<0.001$ &  $0.62$  \\ 
            \hline

            Environment*Day & $4$ & $56$ & $0.70$ & $0.59$ &  $0.05$ 
            &  $4$  & $60$  & $5.98$  &  \cellcolor{lightgray}$<0.001$   & $0.28$   
            &  $4$  & $60$  & $10.32$  &  \cellcolor{lightgray}$<0.001$   & $0.41$  
            &  $4$  & $60$  & $12.98$  &  \cellcolor{lightgray}$<0.001$   & $0.46$  \\ 

            Environment*Time & $2$ & $28$ & $5.70$ & \cellcolor{lightgray}$0.008$ &  $0.29$
            & $2$   & $30$  & $4.07$  &  \cellcolor{lightgray}$0.03$   & $0.21$   
            & $2$   & $30$  & $19.06$  &  \cellcolor{lightgray}$<0.001$   & $0.56$  
            & $2$   & $30$  & $27.10$  &  \cellcolor{lightgray}$<0.001$   & $0.64$  \\ 
            \hline
            
            Day*Time & $8$ & $112$ & $0.72$ & $0.67$ &  $0.05$ 
            & $8$ & $120$ & $5.62$ & \cellcolor{lightgray}$<0.001$ &  $0.27$ 
            & $8$ & $120$ & $1.20$ & $0.30$ &  $0.07$ 
            & $8$ & $120$ & $1.70$ & $1.12$ &  $0.10$  \\ 
            \hline 
            
            Environment*Day *Time & $8$ & $112$ & $0.84$ & $0.57$ &  $0.06$ 
            &  $8$  & $120$  & $3.54$  & \cellcolor{lightgray}$0.001$   & $0.19$   
            &  $8$  & $120$  & $2.21$  & $0.03$   & $0.13$   
            &  $8$  & $120$  & $1.28$  & $0.26$   & $0.08$  \\ 
            
            \hline 

        \end{tabular}

    \label{tab:resultsSTAISSQVF}
    \vspace{-0.3cm}
\end{table*}

\begin{table*}[t]
    \centering 
    \tiny
    \setlength{\tabcolsep}{5pt}
        \caption{RM-ANOVA results for number of breaks, duration of breaks, heart rate and typing speed. d$f_1$ = d$f_{effect}$ and d$f_2$ = d$f_{error}$.}
        \begin{tabular}{|c||c|c|c|c|c||c|c|c|c|c||c|c|c|c|c||c|c|c|c|c|}
            \hline 
            &\multicolumn{5}{|c|}{Number of Breaks} &\multicolumn{5}{|c|}{Break Duration} &\multicolumn{5}{|c|}{Heart Rate} &\multicolumn{5}{|c|}{Typing Speed}  \\
            \hline 
            & d$f_{1}$ & d$f_{2}$ & F & p &  $\eta^2_p$  
            & d$f_{1}$ & d$f_{2}$ & F & p &  $\eta^2_p$
            & d$f_{1}$ & d$f_{2}$ & F & p &  $\eta^2_p$
            & d$f_{1}$ & d$f_{2}$ & F & p &  $\eta^2_p$
             \\ 
            \hline 

            Environment & $1$   & $13$  & $0.87$  &    $0.37$ &  $0.06$  
            & $1$ & $13$ & $12.27$ & \cellcolor{lightgray}$0.004$ &  $0.49$
            & $1$   & $9$  & $0.08$  &    $0.78$ &  $0.009$
            & $1$ & $15$  & $8.46$  & \cellcolor{lightgray}$0.011$  & $0.36$
             \\ 

            \hline 
            
            Day & $4$ & $52$ & $2.7$ & $0.04$ &  $0.17$  
            & $4$ & $52$ & $2.18$ & $0.08$ &  $0.14$
            & d$f_{1}$ & d$f_{2}$ & F & p &  $\eta^2_p$
            & $1$ & $15$  & $8.38$  & \cellcolor{lightgray}$0.011$  & $0.36$
             \\ 
            \hline 
            
            Time & $1$ & $13$ & $0.46$ & $0.51$ &  $0.03$  
            & $1$ & $13$ & $0.04$ & $0.84$ &  $0.003$
            & d$f_{1}$ & d$f_{2}$ & F & p &  $\eta^2_p$
            &  &  &  &  &  
             \\ 
            \hline 

            Environment*Day &  $4$  & $52$  & $1.57$  & $0.20$   & $0.11$   
            & $4$ & $52$ & $2.01$ & $0.11$ &  $0.13$
            & $4$  & $36$  & $1.77$  & $0.16$   & $0.17$
            & $1$  & $15$   & $0.31$   & $0.59$   & $0.02$
             \\ 

            Environment*Time & $1$   & $13$  & $0.71$  & $0.41$   & $0.05$     
            & $1$ & $13$ &  $<0.001$ & $0.98$ &  $.0 
            &   &   &   &   &   
            &   &   &   &   &   
             \\ 
             
            Day*Time & $4$   & $52$  & $0.88$  & $0.48$   & $0.06$     
            & $4$ & $52$ & $0.89$ & $0.48$ &  $0.06$
            &   &   &   &   &   
            &   &   &   &   &   
             \\  
            
            \hline
            
            Environment*Day *Time &  $4$  & $52$  & $2.30$  & $0.07$   & $0.15$   
            & $4$ & $52$ & $1.62$ & $1.18$ &  $0.11$
            &   &   &  &  &   
            &   &   &   &   &   
             \\ 
            
            \hline 

        \end{tabular}

    \label{tab:resultsBreaks}
    \vspace{-0.3cm}
\end{table*}



\typeout{}
\bibliography{template}